\documentclass[%
 reprint,amsmath,amssymb,aps, nofootinbib
]{revtex4-2}

\usepackage{graphicx}
\usepackage[dvipsnames]{xcolor}
\usepackage[colorlinks=true,
    linkcolor=blue,
    filecolor=blue,
    citecolor=blue,      
    urlcolor=blue,]{hyperref}
    
\usepackage[per-mode=symbol-or-fraction, output-decimal-marker={.}, exponent-product = { \cdot }]{siunitx} 

\newcolumntype{R}[1]{>{\raggedleft\let\newline\\\arraybackslash\hspace{0pt}}m{#1}}
\usepackage{xhfill}

\usepackage{cleveref}
\Crefname{figure}{Fig.}{Figs.}

\usepackage{tensor}

\begin{document}

\title {Incorporating static intersite correlation effects in vanadium dioxide through DFT+\textit{V}}

\author{Lea Haas}
\thanks{These authors contributed equally to this work.}
\author{Peter Mlkvik}
\thanks{These authors contributed equally to this work.}
\email{peter.mlkvik@mat.ethz.ch}
\author{Nicola A. Spaldin}
\author{Claude Ederer}
\affiliation{Materials Theory, Department of Materials, ETH Z\"{u}rich, Wolfgang-Pauli-Strasse 27, 8093 Z\"{u}rich, Switzerland}

\date{\today}

\begin{abstract}

We analyze the effects on the structural and electronic properties of vanadium dioxide (VO$_2$) of adding an empirical inter-atomic potential within the density-functional theory$+V$ (DFT$+V$) framework. We use the DFT$+V$ machinery founded on the extended Hubbard model to apply an empirical self-energy correction between nearest-neighbor vanadium atoms in both rutile and monoclinic phases, and for a set of structures interpolating between these two cases. We observe that imposing an explicit intersite interaction $V$ along the vanadium-vanadium chains enhances the characteristic bonding-antibonding splitting of the relevant bands in the monoclinic phase, thus favoring electronic dimerization and the formation of a band gap. We then explore the effect of $V$ on the structural properties and the relative energies of the two phases, finding an insulating global energy minimum for the monoclinic phase, consistent with experimental observations. With increasing $V$, this minimum becomes deeper relative to the rutile structure, and the transition from the metallic to the insulating state becomes sharper. We also analyze the effect of applying the $+V$ correction either to all or only to selected vanadium-vanadium pairs, and both in the monoclinic as well as in the metallic rutile phase. Our results suggest that  DFT$+V$ can indeed serve as a computationally inexpensive unbiased way of modeling VO$_2$ which is well suited for studies that, e.g., require large system sizes.
\end{abstract}

\maketitle


\section{Introduction}
\label{sec:intro}

\begin{figure}[!t]
    \centering
    \includegraphics[width=\linewidth]{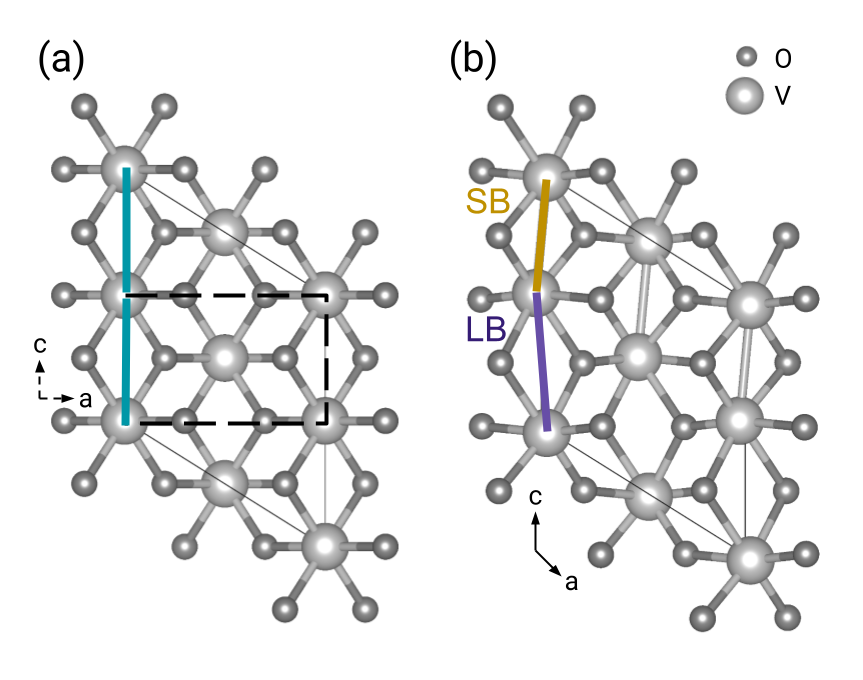}
    \caption{(a) The R structure of VO$_2$ depicted in the M1 unit cell, indicating the orientation of the R lattice vectors. The primitive unit cell of the R structure is also indicated by the black dashed lines. Two vanadium--vanadium bonds of equal length are shown in turquoise. (b) M1 structure of VO$_2$ with the vanadium--vanadium short bond (SB) and long bond (LB) shown in yellow and purple, respectively. Vanadium (oxygen) atoms shown in light (dark) grey.}
    \label{fig:vo2_structure}
\end{figure}

Vanadium dioxide (VO$_2$) is a prototypical material exhibiting a metal-insulator transition (MIT) coupled with a structural transition. Its first-order MIT is accompanied by a structural distortion between the high-symmetry, high-temperature rutile (R) phase and a low-symmetry, low-temperature monoclinic (M1) phase~\cite{Eyert:2002}. The M1 phase is characterized by the formation of zig-zagging dimers along chains of vanadium atoms parallel to the $c$ direction  (see Fig.~\ref{fig:vo2_structure}), which leads to a doubling of the unit cell reminiscent of a Peierls distortion. Since on the other hand the  V$^{4+}$ cations in VO$_2$ are formally occupied by a single electron in the rather localized $d$ orbitals, the insulating M1 phase of VO$_2$ has been widely described to arise from an interplay between Mott-Hubbard and Peierls-like physics.

Already in the 1970s, Goodenough~\cite{Goodenough:1971} postulated that the gap formation in the M1 insulating phase can be explained by a bonding-antibonding splitting of the lowest-lying vanadium~$t_{2g}$ orbital, the $a_{1g}$, with lobes pointing towards the nearest-neighbor vanadium atoms along $c$, and the increase in energy of the other two $t_{2g}$ orbitals (often labeled $e_g^\pi$\footnote{We acknowledge that this is strictly not the correct symmetry label for the V states in both the R and M1 phases of VO$_2$. All three $t_{2g}$ levels are split and hence the two ``$e_g^\pi$'' levels are not degenerate in energy. However, we choose to continue the standard nomenclature as used in other references in the field.}). While there is general agreement on this overall picture, the role of electronic correlations in VO$_2$ has been debated for decades~\cite{Zylbersztejn/Mott:1975, Wentzcovitch/Schulz/Allen:1994, Rice/Launois/Pouget:1994}, with both experimental~\cite{Qazilbash_et_al:2007, Lee_et_al:2018} and theoretical~\cite{Gatti_et_al:2007, Weber_et_al:2012, Brito_et_al:2016, Najera_et_al:2017, Grandi/Amaricci/Fabrizio:2020} studies emphasizing the importance of Mott-Hubbard electronic correlation effects in VO$_2$.

Nowadays, state-of-the-art computational methods such as (cluster) dynamical mean-field theory (DMFT) can capture the different phases of VO$_2$ quite well~\cite{Biermann_et_al:2005, Brito_et_al:2016, Weber_et_al:2020}. However, while a two-site cluster approach needs to be used for the M1 phase, the R phase is typically treated with standard single-site DMFT, to not incorporate a possible bias by grouping specific pairs of vanadium atoms into fixed clusters. This distinct treatment for the R and M1 phases makes it difficult to systematically vary the degree of dimerization between the two phases or to treat dopants and substitutional atoms. Additionally, the high complexity and computational cost of the cluster DMFT method restricts the maximum system size that can be afforded. However, such studies would be highly desirable in view of the many proposed advanced technological devices that exploit the possibility of tuning the MIT in VO$_2$ via doping, strain, or heterostructuring, for example in the fields of neuromimetic circuits~\cite{Andrews_et_al:2019, Yi_et_al:2018, Lin/Guha/Ramanathan:2018}, electrical switches~\cite{Yang/Ko/Ramanathan:2011, Vitale_et_al:2016}, or smart windows~\cite{Cui_et_al:2018}.

In this work, we address these two deficiencies by drawing on previous studies that emphasized the importance of inter-atomic effects in VO$_2$. In Biermann~{\it et al.}~\cite{Biermann_et_al:2005}, the authors performed the first DFT + cluster DMFT study on VO$_2$, demonstrating the importance of the intra-dimer self-energy. Follow up works~\cite{Tomczak/Biermann:2007, Tomczak/Aryasetiawan/Biermann:2008, Tomczak/Biermann:2009} then found that modeling the almost frequency-independent inter-atomic self-energy as static yields results that are comparable to full cluster DMFT calculations. Drawing on these previous works, Belozerov~{\it et al.}~\cite{Belozerov_et_al:2012} performed so-called DFT$+V$ + single-site DMFT calculations on VO$_2$, using an empirically chosen inter-atomic potential $V$ to enhance the bonding-antibonding splitting of the $a_{1g}$ orbitals in the M1 phase. This was then combined with single-site DMFT to treat on-site correlations. Recently, the static inter-atomic self-energy approach has also been used in conjunction with single-site DMFT applied to the $e_g^\pi$ orbitals in hydrogen-doped VO$_2$~\cite{Kim_et_al:2022}. Finally, He and Millis~\cite{He/Millis:2016} studied non-equilibrium excitations in VO$_2$ to describe the photoinduced MIT also employing an empirical inter-atomic $V$ term. The authors used DFT$+U$$+V$~\cite{LeiriaCampoJr/Cococcioni:2010} to account for both on-site and inter-atomic interactions, and were able to describe the experimentally reported photo-induced metallic M1 phase.

Here, we build on this previous research and show that a ``bias-free'' description can be obtained by utilizing DFT$+V$ and applying the intersite correction between all neighboring vanadium atoms along $c$ and not only between the dimerized pairs. We apply this method to both the non-dimerized R phase and the dimerized M1 phase, thereby treating both phases on an equal footing. This allows us to study the evolution of the MIT with increasing structural distortion between the main VO$_2$ phases in a simple and bias-free way with low computational cost. 

We follow the DFT$+U$$+V$ approach outlined by Campo and Coccocioni~\cite{LeiriaCampoJr/Cococcioni:2010} using $U=0$ and applying the inter-atomic potential $V\neq0$ between neighboring vanadium atoms along $c$. 
We study the effect of $V$ on structural properties, crucially showing that both structural dimerization and the addition of a $V$ favor larger bonding anti-bonding splitting of the bands. An important finding is that, in contrast to conventional DFT, the DFT$+V$ approach correctly obtains a sizable insulating gap in the M1 phase as well as the correct energetic ordering between the M1 and R phases. 
To better understand the effects of $V$ on the electronic structure, we also compare our results to a simple one-dimensional tight-binding model. 

Thus, our results provide insights into the effects of inter-atomic electron correlations in VO$_2$, that are complementary to that of previous works, and demonstrate the utility of the DFT$+V$ approach for describing complex correlated systems, thereby providing a good compromise between accuracy and computational efficiency. 

\section{Methodology}

\subsection{DFT+\textit{V} in the context of this work}

We first elaborate on the use of DFT$+V$ within the scope of this work. In the DFT($+U$)$+V$ approach, the standard DFT functional is supplemented by terms based on the extended Hubbard model~\cite{Hubbard:1963}:
\begin{equation}
\begin{split}
\label{eq:ext_hubb}
\hat{H} = \sum_{\langle ij \rangle \sigma} t_{ij} (\hat{c}_{i \sigma}^\dag \hat{c}_{j \sigma} + \text{h.c.}) + \sum_{i} U_i \hat{n}_{i \uparrow} \hat{n}_{i  \downarrow} \\
+ \sum_{\langle ij \rangle\sigma\sigma'} V_{ij} \hat{n}_{i\sigma} \hat{n}_{j\sigma'} \ ,
\end{split}
\end{equation}
where $i$ and $j$ refer to the atomic sites with creation (annihilation) operators $\hat{c}_{i \sigma}^{\dag}$ ($\hat{c}_{i \sigma}$) for spin $\sigma$, $\hat{n}_{i \sigma}$ is the number operator for electrons on site $i$ with spin $\sigma$, $t_{ij}$ is the hopping amplitude between sites $i$ and $j$, and $U_{i}$ and $V_{ij}$ are the on-site and inter-atomic Coulomb interactions respectively. 

The DFT+$V$ approach corresponds to the use of a mean-field approximation for the inter-atomic interaction in the last term of the Hamiltonian of Eq.~(\ref{eq:ext_hubb}) (see, e.g., Ref.~\cite{Belozerov_et_al:2012}):
\begin{equation}
    \begin{split}
    \hat{n}_{i\sigma} \hat{n}_{j\sigma'}
    & = \hat{c}_{i\sigma}^{\dag}\hat{c}_{i\sigma} \hat{c}_{j\sigma'}^{\dag}\hat{c}_{j\sigma'} \\
 \rightarrow \quad & \hat{c}_{i\sigma}^{\dag}\hat{c}_{i\sigma} \langle \hat{c}_{j\sigma'}^{\dag}\hat{c}_{j\sigma'} \rangle + \hat{c}_{j\sigma'}^{\dag}\hat{c}_{j\sigma'} \langle \hat{c}_{i\sigma}^{\dag}\hat{c}_{i\sigma} \rangle \\
    & - \hat{c}_{j\sigma'}^{\dag}\hat{c}_{i\sigma} \langle \hat{c}_{i\sigma}^{\dag}\hat{c}_{j\sigma'} \rangle - \hat{c}_{i\sigma}^{\dag}\hat{c}_{j\sigma'} \langle \hat{c}_{j\sigma'}^{\dag}\hat{c}_{i\sigma} \rangle  \ .
    \end{split}
\end{equation}
With this, the inter-atomic interaction term can be written as:
\begin{equation}
\label{eq:intersite}
 \sum_{\langle ij \rangle \sigma\sigma'} V_{ij} \hat{c}^\dag_{i\sigma} \hat{c}_{i\sigma} \langle \hat{c}^\dag_{j\sigma'} \hat{c}_{j\sigma'} \rangle - \sum_{\langle ij \rangle \sigma} V_{ij} \hat{c}^{\dag}_{i\sigma} \hat{c}_{j\sigma} \langle c^{\dag}_{j\sigma} c_{i\sigma} \rangle \ .
\end{equation}
The first term in Eq.~(\ref{eq:intersite}) represents the normal intersite Hartree interaction, which in the framework of the DFT+$V$ method is accounted for through the standard DFT energy functional. The second term in Eq.~(\ref{eq:intersite}) represents an intersite exchange interaction. Within DFT+$V$, a term of this form is explicitly added to the Kohn-Sham potential to account for intersite effects that are not well described by standard DFT functionals~\cite{LeiriaCampoJr/Cococcioni:2010}. One can see that this term leads to an effective renormalization of the intersite hopping according to 
\begin{equation}
    \label{eq:renorm}
\tilde{t}_{ij}^\sigma = t_{ij} - V_{ij} n_{ji}^\sigma \ , 
\end{equation}
where the additional contribution is given by the interaction strength $V_{ij}$ weighted by the \emph{inter-atomic occupation} $n_{ji}^\sigma = \langle \hat{c}_{j\sigma}^{\dag}\hat{c}_{i\sigma} \rangle$. We note that in Eq.~(\ref{eq:intersite}) we have assumed that the intersite occupation matrix is spin-diagonal, which is certainly true for all systems without non-collinear magnetic order or spin-orbit coupling. In the following, we do not consider any spin polarization and denote the intersite occupation as $n_\times$ -- ``cross'' to denote the cross term of the occupation matrix. Consistent with the crystal structure of VO$_2$, and guided by previous works demonstrating the importance of the corresponding intersite effects~\cite{Tomczak/Biermann:2007, Tomczak/Aryasetiawan/Biermann:2008, Tomczak/Biermann:2009, Belozerov_et_al:2012, He/Millis:2016}, we apply $V_{ij} \neq 0$ between neighboring vanadium sites along the $c$ axis, affecting the hopping between the vanadium atoms along the nearest-neighbor chains, while treating all other neighbors only at the plain DFT level. 

Throughout this work, we treat the intersite interaction $V$ as an adjustable parameter, which we apply either to all or to only selected neighboring vanadium-vanadium pairs along the $c$ direction, thereby modifying the respective 
effective hopping amplitudes. In particular, we analyze how the application of $V$ affects the non-dimerized rutile phase, and, for the dimerized M1 phase, we compare the effect of applying the inter-atomic term on all vanadium-vanadium neighbor pairs along $c$ with the cases where we consider a $V$ only for the short bond (SB) or long bond (LB) pairs, respectively ({\it cf.} \Cref{fig:vo2_structure}). In this way, we systematically monitor the effects of varying the size and also the sign of $V$ on the different phases of VO$_2$, i.e., both in the dimerized M1 and non-dimerized R structures, as well as for structures obtained by interpolating between these two. 

We point out that, in line with previous studies~\cite{Tomczak/Biermann:2007, Tomczak/Aryasetiawan/Biermann:2008, Tomczak/Biermann:2009, Belozerov_et_al:2012, He/Millis:2016}, the parameter $V$ in this work should be viewed as an effective inter-atomic self-energy correction, representing inter-atomic correlations that ultimately result from local interactions, and not as an actual inter-atomic Coulomb interaction. In fact, we estimated the strength of the nearest neighbor vanadium-vanadium Coulomb interaction using the linear response/density functional perturbation theory-based approach of Refs.~\cite{Timrov/Marzari/Cococcioni:2018, Timrov/Marzari/Cococcioni:2022} and obtained only a negligible inter-atomic interaction of $\approx 0.01$\,eV. In contrast, the static component of the inter-atomic self energy used in earlier works corresponds to $V=2$\,eV~\cite{Belozerov_et_al:2012} or $V=1$\,eV~\cite{He/Millis:2016, Kim_et_al:2022}, similar to the values we use in our work. 

To interpret our DFT+$V$ results, we also relate them to a simple one-dimensional tight-binding model, corresponding to the vanadium $a_{1g}$ orbitals along the nearest neighbor chains along $c$, with renormalized hopping amplitudes according to Eq.~(\ref{eq:renorm}). Due to the different vanadium--vanadium distances in the M1 phase -- short bond (SB) and long bond (LB), we consider two distinct hopping amplitudes $t^\text{SB}$ and $t^\text{LB}$, two distinct inter-atomic occupations $n_\times^\text{SB}$ and $n_\times^\text{LB}$, and two distinct interactions $V^\text{SB}$ and $V^\text{LB}$ for the short and long bonds respectively. This can be expressed as:
\begin{equation}
    \hat{H} = \sum_\ell (\tilde{t}^\text{\,SB} \hat{c}_{1, \ell}^{\dag} \tensor{\hat{c}}{_{2, \ell}} +\tilde{t}^\text{\,LB} \hat{c}_{1, \ell+1}^\dag \tensor{\hat{c}}{_{2, \ell}} + \textrm{h.c.}).
\end{equation}
with $\hat{c}_{i, \ell}$ referring to vanadium atom $i \in \{1, 2\}$ in unit cell $\ell$ of the M1 structure and $\hat{c}_{i, \ell+1}$ to the atoms in its neighboring unit cell along $c$. We note that generally the intersite occupations $n_\times^\text{SB/LB}$ depend on $V^\text{SB/LB}$ and, in principle, have to be determined self-consistently. If the resultant $\tilde{t}^\text{\,SB} \neq \tilde{t}^\text{\,LB}$, a band gap opens at the Brillouin zone boundary. The width of this band gap, $\Delta$, and the bandwidth, $w$, of the resulting upper or lower band are given by:
\begin{equation}
 \label{eq:gap}
     \Delta = 2 | \tilde{t}^\text{\,SB} - \tilde{t}^\text{\,LB}| \ ,
\end{equation}
\begin{equation}
 \label{eq:bandwidth}
     w = |\tilde{t}^\text{\,SB} + \tilde{t}^\text{\,LB}| - |\tilde{t}^\text{\,SB} - \tilde{t}^\text{\,LB}| \ .
\end{equation}

\subsection{DFT calculation details}

For most of the calculations throughout this work, we use the unit cell of the lower-symmetry M1 structure~(shown in \Cref{fig:vo2_structure}). This can be viewed as a supercell of the R structure, doubled along the $c$ lattice vector and described using non-orthogonal monoclinic lattice vectors [\Cref{fig:vo2_structure}(a)]. For the structural distortion described in Sec.~\ref{sec:struc_distort}, we use the experimental lattice parameters of the R phase~\cite{McWhan_et_al:1974} also to treat the M1 phase within this cell, thereby neglecting both the monoclinic strain as well as the expansion of the $c$ lattice parameter (and any other volume-related changes) in the M1 phase relative to R [see the comparison of ``Simp.'' (for simplified) versus ``Exp.'' (for experimental) M1 structures in Table~\ref{tab:relax}]. This approximation facilitates a simpler interpolation between the M1 and R phases, since only the internal atomic positions need to be varied. We have verified that this choice has little to no effect on the resultant electronic structure in the scope of our work here. In our full structural relaxations, we then optimize both the internal atomic positions and the corresponding lattice parameters (including the monoclinic strain in the M1 phase). 

We perform DFT$+V$ calculations using the \textsc{Quantum ESPRESSO}~(QE 7.1 and 7.2) package~\cite{Giannozzi_et_al:2009, Giannozzi_et_al:2017} within the generalized gradient approximation using the Perdew-Burke-Ernzerhof~(PBE)~\cite{Perdew/Burke/Ernzerhof:1996} exchange-correlation functional. We use the ultrasoft pseudopotentials from the GBRV library~\cite{Garrity_et_al:2014}, including the $3s$ and $3p$ semi-core states in the valence manifold for the vanadium atoms. We use a plane-wave kinetic energy cut-off of $70$~Ry for the wavefunction and 12$\times$70~Ry for the charge density. We use a $\Gamma$-centered $6\times6\times8$ $k$-point grid, converging total energies to a tolerance of \SI{5e-8}{\electronvolt} between successive iteration steps.


\section{Results and Discussion}

\subsection{Effects of \textit{V} on the DFT results for M1 and R phases}
\label{sec:dftv_effects}

\begin{figure}
    \centering
    \includegraphics[width=\linewidth]{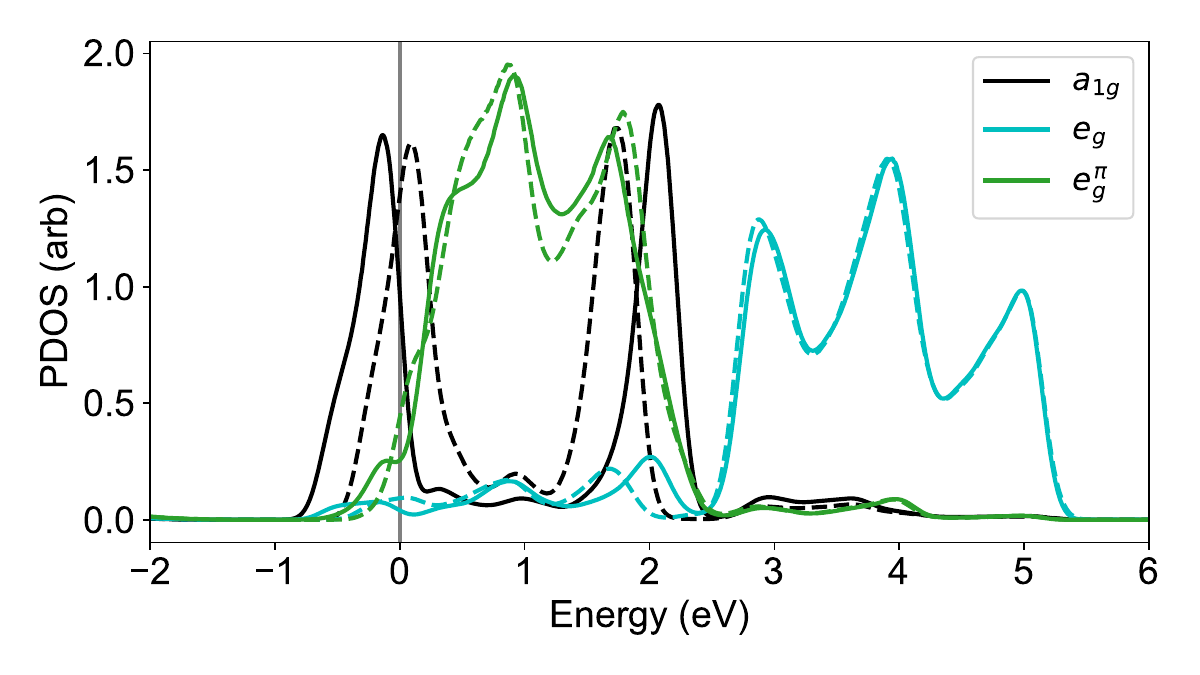}
    \caption{Orbital projected densities of states (PDOS) corresponding to the different sets of vanadium $d$ orbitals in the M1 phase of VO$_2$, obtained for $V=0$\,eV (dashed lines) and $V=1$\,eV (solid lines), applied on all vanadium-vanadium pairs along $c$. The $a_{1g}, e_g^\pi$, and $e_g$ orbitals are shown, respectively, in black, green, and blue. Zero energy is set to the Fermi level of the $V=1$\,eV case, so that the difference in the $a_{1g}$ bonding-antibonding splitting is more apparent.}
    \label{fig:different_orbs}
\end{figure}

Before discussing the results of our structural relaxations, we first explore the effects of a $V$ on the M1 and R phases using their respective experimental lattice parameters given as ``Exp.'' in Table~\ref{tab:relax}. We treat the $V$ as a free parameter, varying it in both positive and negative directions, and also applying it to different vanadium-vanadium pairs. 

We start our analysis with the M1 structure. In \Cref{fig:different_orbs}, we compare orbital-projected densities of states obtained for $V=0$ (the PBE result) and $V=1$\,eV. It can be seen that application of a $V$ mainly affects the $a_{1g}$ orbitals, whereas the other $d$ orbital components remain nearly unaffected (see dashed versus full lines in \Cref{fig:different_orbs}). This is due to the large $a_{1g}$ component ($n_\times^{a_{1g}}\simeq0.3$) of the inter-atomic occupation matrix, $n_\times^\text{SB}$, resulting from the electronic dimerization. In contrast, the inter-atomic elements corresponding to the other orbitals and also the LB pairs remain small ($n_\times^{\text{LB}}\lesssim0.1$). This, in turn, means that the corresponding renormalization of the effective hopping [as described in Eq.~(\ref{eq:renorm})] is negligibly small for the $e_g^\pi$ (which form the remainder of the $t_{2g}$ manifold in an ideal octahedron) and the $e_g^\sigma$ orbitals. In the following we therefore focus on the analysis of the $a_{1g}$ projected density of states (PDOS) only.

\begin{figure}
    \centering
    \includegraphics[width=1\linewidth]{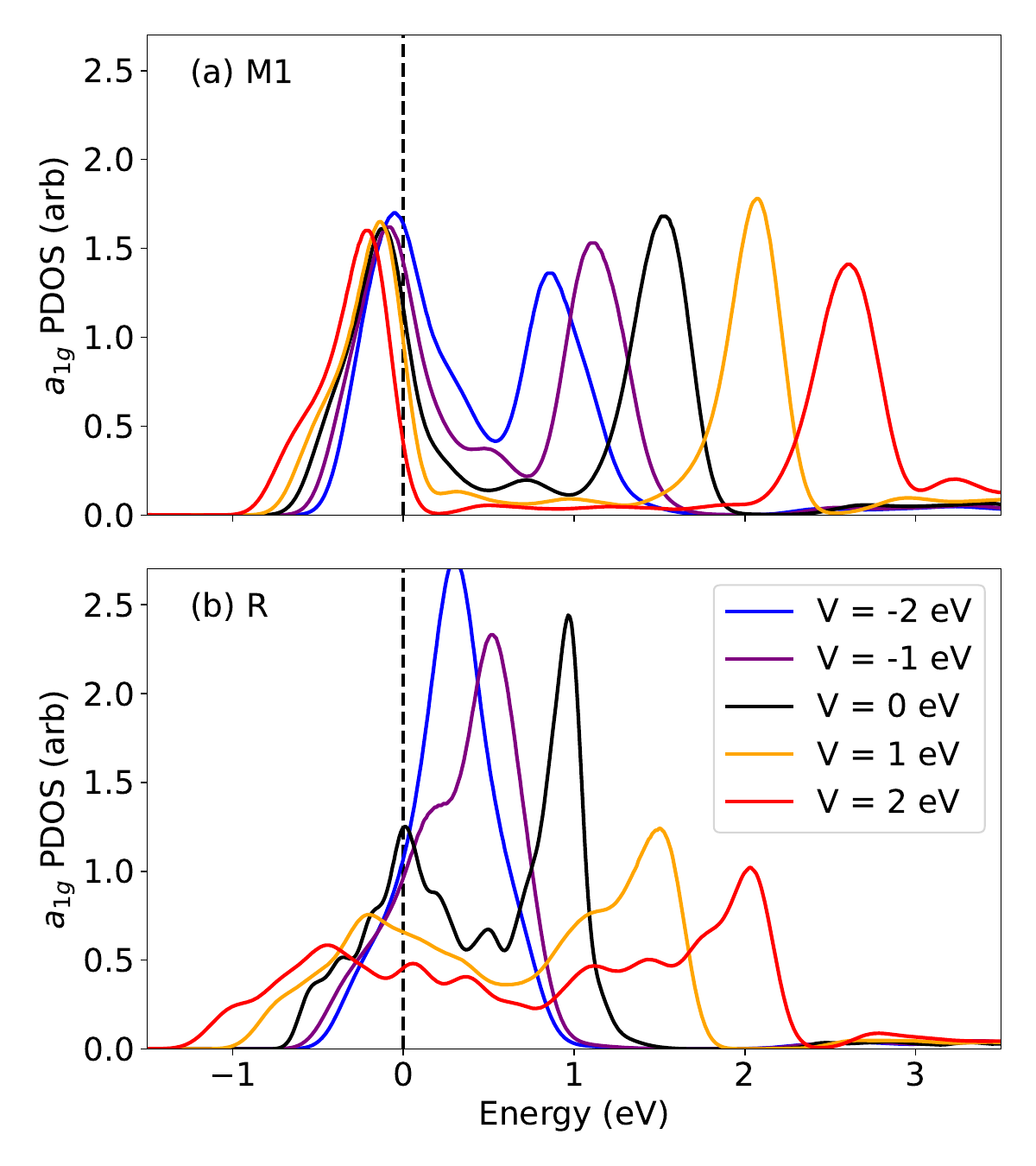}
    \caption{Vanadium $a_{1g}$-orbital PDOS for different positive and negative values of $V$ (shown in red and blue color gradient respectively, with $V=0$ in black) applied on both bonds for (a) M1 and (b) R structures of VO$_2$. Fermi level at 0\,eV shown by the dashed line.}
    \label{fig:dx2-y2_pdos}
\end{figure}

In \Cref{fig:dx2-y2_pdos}(a), we further highlight the effect of a $V$ on the $a_{1g}$ bonding-antibonding splitting, which varies approximately linearly with $V$, for both positive and negative values of $V$. This, in turn, leads to a band gap linearly increasing with $V$, since the width of the bonding and antibonding peaks remain almost constant. We can compare this behavior with the simple one-dimensional tight binding model, specifically Eqs.~(\ref{eq:gap}) and (\ref{eq:bandwidth}). In the M1 phase, we find that $n_\times^\text{LB} \ll n_\times^\text{SB}$ and so the tight binding model yields a linearly changing $\Delta$ with a constant $w$ for a changing $V$, mirroring exactly the DFT results [\Cref{fig:dx2-y2_pdos}(a)]. This trend persists as long as $n_\times$ is approximately constant. We note that the bare hopping between $a_{1g}$ orbitals on neighboring vanadium sites is negative, i.e., $t^\text{SB/LB}<0$, and thus according to Eq.~(\ref{eq:renorm}) a positive $V$ (for positive $n_\times$) leads to an increase in the absolute value of the effective hopping amplitude.

\begin{figure}
    \centering
    \includegraphics[width=1\linewidth]{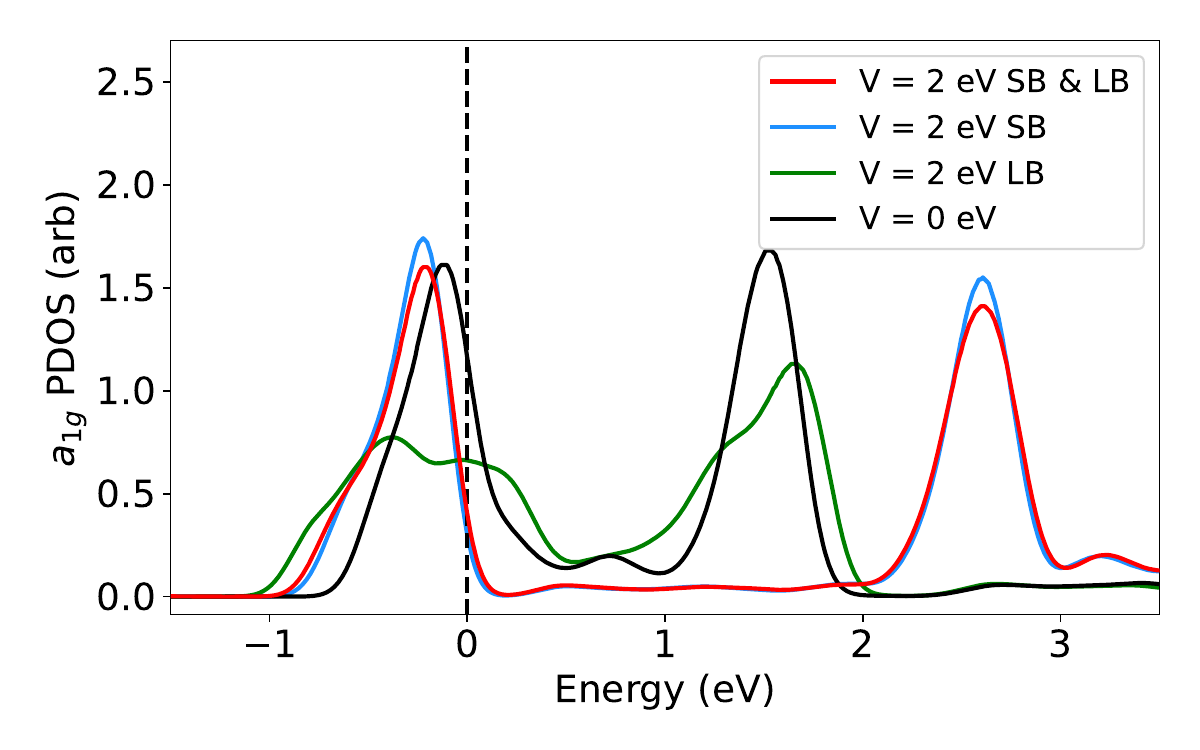}
    \caption{Vanadium~$a_{1g}$-orbital PDOS of the M1 phase, obtained by applying an inter-atomic potential of $V=2$\,eV to both SB and LB (red line) or to either SB (cyan line) or LB (magenta line) vanadium-vanadium pairs. The black line shows the case with $V=0$\,eV. Fermi level at 0\,eV shown by the dashed line.}
    \label{fig:diffv_pdos}
\end{figure}

Next, we analyze how applying a $V$ only on either the SB or LB vanadium-vanadium pairs affects the electronic structure of the M1 phase. The results, depicted in \Cref{fig:diffv_pdos}, show that applying the $V$ to all vanadium-vanadium pairs is nearly identical in behavior to only applying it to the SB pairs (compare red and blue lines in \Cref{fig:diffv_pdos}). This is due to the fact that we are in a regime with $n_\times^\text{LB} \approx 0$, so that the $V$ on the LB does not change the effective hopping, giving the same bandwidth and band gap as in the SB-only case.

Testing the effects of applying a $V$ only on the LB in the M1 phase, we observe a broadening of the PDOS for positive $V$ values (note the difference between the green and black lines in \Cref{fig:diffv_pdos}). This is because $n_\times^\text{LB}$ is increased, favoring dimerization on the LB, working against the existing SB dimerization.

\begin{figure*}
    \centering
    \includegraphics[width=1\textwidth]{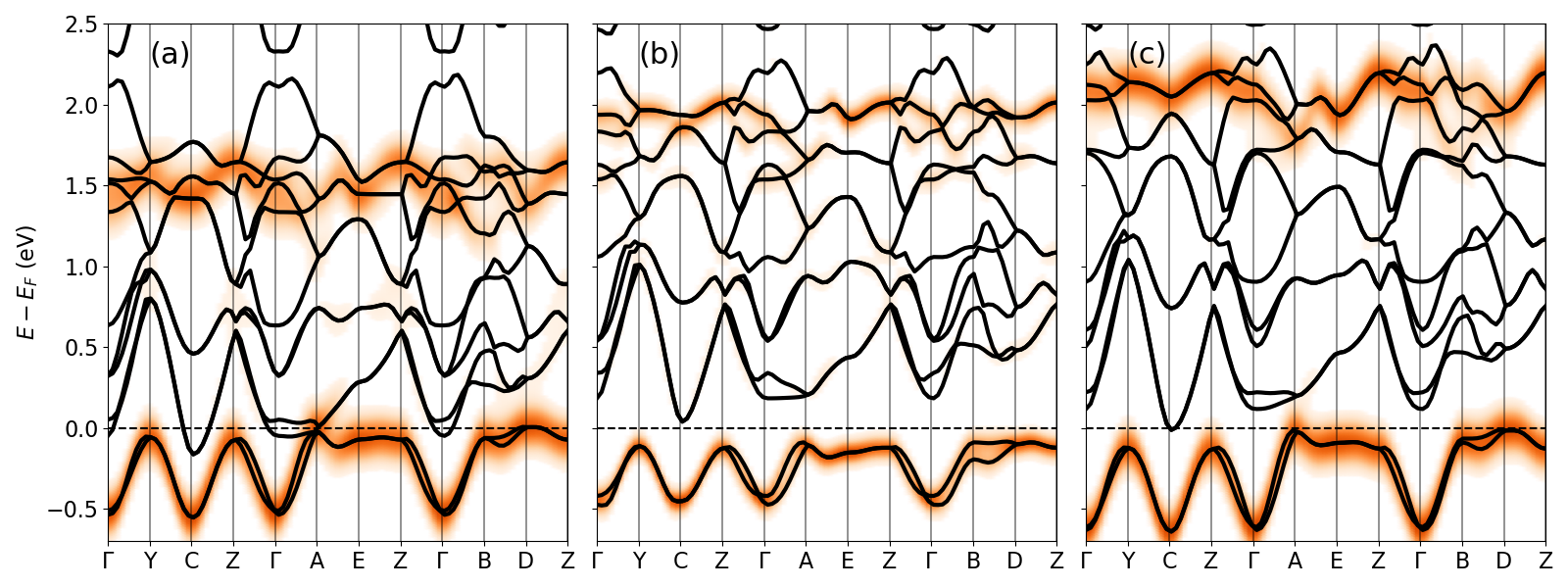}
    \caption{Band structures of the M1 phase with the projection on the $a_{1g}$ orbital highlighted. (a) Experimental (``Exp.'') structure at $V=0$. (b) Relaxed structure at $V=0$. (c) Experimental structure with $V=1$\,eV. The Fermi energy at 0\,eV shown in dashed line.}
    \label{fig:bandstr_relaxed_plusv}
\end{figure*}

Finally, the changes in the PDOS when applying a $V$ on all vanadium-vanadium pairs are also reflected in the band structure, causing a separation of bands and a band gap formation in the M1 phase [see the band structure for $V=0$ and $V=1$\,eV in \Cref{fig:bandstr_relaxed_plusv}(a, c)]. The increased bonding-antibonding splitting of the $a_{1g}$ orbital as indicated by the highlighted bands leads to a gap-opening reminiscent of the Goodenough picture described in Sec.~\ref{sec:intro}.

The observation that applying a $V$ to all vanadium-vanadium pairs gives equivalent results compared to applying it only to the SB pairs is an important result, since this allows for a bias-free way of treating the system, i.e. without a pre-pattering of vanadium dimers. In order to fully describe the distortion in VO$_2$, we now evaluate the effects of applying a $V$ onto the chains with equidistant vanadium atoms present in the R phase.

Indeed, for the R phase [\Cref{fig:dx2-y2_pdos}(b)], we see a strikingly different behavior compared to the M1 case [\Cref{fig:dx2-y2_pdos}(a)]. For positive values of $V$, the $a_{1g}$ density of states broadens without opening a gap, while for negative values of $V$, we observe a narrowing of the PDOS into a single sharp peak. To understand this result in terms of a simple one-dimensional tight-binding model, we set $n_\times^\text{SB}=n_\times^\text{LB}$, $V^\text{SB}=V^\text{LB}$, and $t^\text{SB}=t^\text{LB}$. Since the two effective hopping amplitudes are identical, there is no gap opening [{\it cf.} Eq.~(\ref{eq:gap})]. Instead, varying the $V$ at a constant $n_\times$ simply changes the overall hopping and thus the bandwidth linearly [Eq.~(\ref{eq:bandwidth})], consistent with the changes in the PDOS in \Cref{fig:diffv_pdos}(b).

From the results presented in this section, it thus follows that for $V\gtrsim 1$\,eV one obtains a  metallic R and an insulating M1 phase, in agreement with the measured behavior of VO$_2$, and without a pre-selecting of specific vanadium-vanadium pairs to form dimers. 

\begin{table*}
\caption{Experimental (``Exp.''), simplified (``Simp.''), and other calculated lattice parameters and internal coordinates for both R (``Exp.'' from Ref.~\cite{McWhan_et_al:1974} at 360~K) and M1 (``Exp.'' from Ref.~\cite{Longo_et_al:1970} at 298~K) phases as calculated within DFT$+V$ at different values of $V$ and structural relaxation conditions. Coordinate system corresponds to that defined in Fig.~\ref{fig:vo2_structure}.}
\begin{tabular}{p{0.7   cm}p{2.5cm}|*{8}{R{1.3cm}}}
\hline
\hline
&                       & \multicolumn{1}{c}{Exp.} & \multicolumn{1}{c}{Simp.} & \multicolumn{3}{c}{Internally relaxed} & \multicolumn{3}{c}{Fully relaxed} \\ \hline
& $V$ (eV)              & \multicolumn{1}{c}{0} & \multicolumn{1}{c}{0} & \multicolumn{1}{c}{0} & \multicolumn{1}{c}{1} & \multicolumn{1}{c}{2} & \multicolumn{1}{c}{0} & \multicolumn{1}{c}{1} & \multicolumn{1}{c}{2}  \\ \hline
R & $a$ (\r{A}) & 4.555                  & \multicolumn{4}{r}{\xrfill{0.1pt}}       & 4.622       & 4.627       & 4.630      \\
& $c$ (\r{A}) & 2.851                  & \multicolumn{4}{r}{\xrfill{0.1pt}}       & 2.772    & 2.767       & 2.765      \\[0.3cm]
M1 & $a$ (\r{A})         & 5.383                   & 5.374                    &        \multicolumn{3}{r}{\xrfill{0.1pt}}       & 5.426       & 5.437       & 5.444      \\
& $b$ (\r{A})         & 4.538                   & 4.555                    &        \multicolumn{3}{r}{\xrfill{0.1pt}}       & 4.597   & 4.600       & 4.606      \\
& $c$ (\r{A})         & 5.752                   & 5.703                    &        \multicolumn{3}{r}{\xrfill{0.1pt}}       & 5.650   & 5.646       & 5.640      \\
& $\beta$ ($^{\circ}$)  & 122.6                   & 122.0                    &        \multicolumn{3}{r}{\xrfill{0.1pt}}       & 121.8     & 121.8       & 121.8      \\
& $r_{\text{SB}}$                   & 0.479                       & 0.479                     & 0.464      & 0.462   & 0.462     & 0.467       & 0.466       & 0.466      \\
\hline
\hline
\end{tabular}
\label{tab:relax}
\end{table*}

\subsection{Effects of \textit{V} on the DFT results for \texorpdfstring{VO$_2$}{VO2} structure}
\label{sec:struc_relax}

Next, we perform structural relaxations for both the R and M1 structures using a range of values for $V$. The results are summarized in Table~\ref{tab:relax}. For the R phase we show the two distinct lattice parameters - $a$ within the rutile basal plane, and $c$ along the dimerization axis [see \Cref{fig:vo2_structure}(a)]. For the M1 phase, we present all lattice parameters ($a$, $b$, and $c$, the latter along the dimerization axis), the monoclinic angle ($\beta$), and the ratio of the SB distance projected onto the $c$ lattice parameter relative to the magnitude of the $c$ lattice parameter ($r_\text{SB}$). First, we describe the effect of full relaxations at three different values of $V$ and then we show that only relaxing the atomic coordinates yields very similar results.

In the R phase, we observe a change of the lattice parameters that corresponds to the typical overestimation of the unit cell volume using the PBE functional (Table~\ref{tab:relax}). However, we find that this larger cell volume is accommodated in an anisotropic way through a larger $a$ lattice parameter, while the $c$ lattice parameter is actually smaller than the experimental value, leading to shorter distances between vanadium atoms along the chain direction. We observe very little change in the internal parameters. Applying an inter-atomic $V$ has only a negligible effect on the structural relaxation, manifesting itself only in the third digits in Table~\ref{tab:relax} for both internal and full relaxations. We observe no band gap opening in the R phase.

In the relaxed M1 phase, we first note that, already for $V=0$, we obtain a band gap opening as shown in \Cref{fig:bandstr_relaxed_plusv}(b). This is caused by a change in the internal coordinates that leads to a shortening of the dimer length (quantified through $r_{\text{SB}}$ in Table~\ref{tab:relax}), and hence, in turn, an enhancement of the bonding-antibonding splitting. Notably, this is very similar to what we observe without any relaxation but with the addition of $V$ -- all three band structures are shown in \Cref{fig:bandstr_relaxed_plusv}. 

The overall behavior of the lattice parameters here is analogous to that of the R phase, where with relaxations, the lattice parameter along the dimer direction ($c$) shortens while the other two lattice parameters increase in magnitude. Similarly to the R phase, the addition of $V$ has very little effect on all relaxations. For the internal relaxations in the M1 phase, we use a simplified structure (``Simp''. in Table~\ref{tab:relax}) formed by doubling the R primitive structure and considering it within a monoclinic cell as opposed to the similar M1 experimental structure. Internal relaxations yield similar results to full relaxations, also causing a band gap opening.

\subsection{Effects of \textit{V} on the DFT results across a structural distortion}
\label{sec:struc_distort}

Finally, we study the evolution of VO$_2$ when varying the structure continuously between the R and M1 phases in order to better understand how the metal-insulator transition and how (or whether) the energy landscape changes under the application of a $V$. We do so by varying only the internal atomic positions while keeping the lattice vectors fixed, thereby interpolating between the simplified M1 structure (``Simp.'' in Table~\ref{tab:relax}) and the R structure embedded in the monoclinic cell. We parameterize the degree of the structural distortion by a scalar amplitude, $d$, where $d=0$ corresponds to the undistorted experimental R structure and $d=1.0$ corresponds to the simplified M1 structure (``Simp.''). This simplified treatment of both the structure itself and the structural distortion is justified by the results of the previous section where we showed that for our work we observe no significant differences between full relaxations and internal relaxations under different $V$ values. 

\begin{figure}
    \centering
    \includegraphics[width=0.45\textwidth]{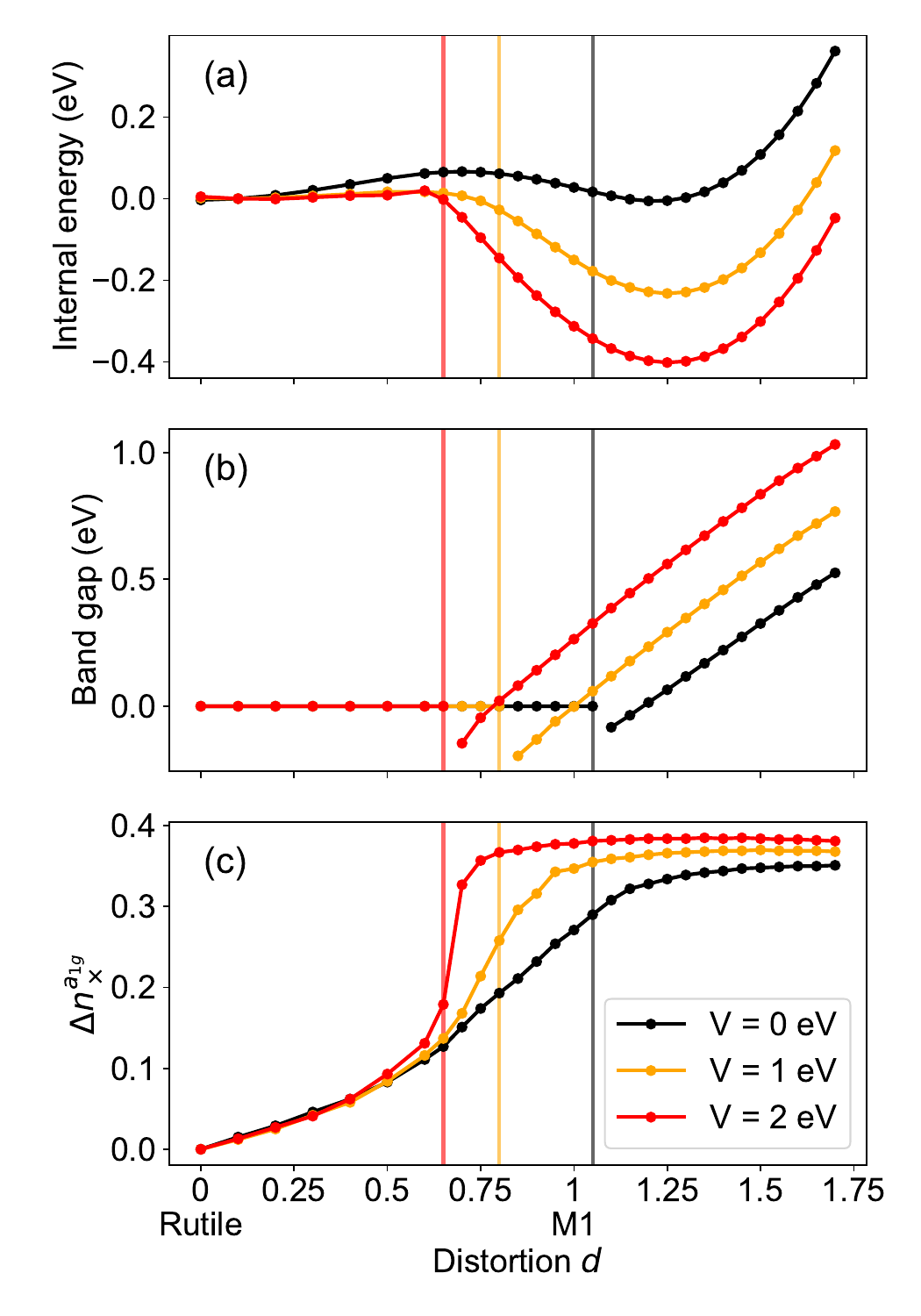}
    \caption{Properties along the distortion $d$ interpolating between and extrapolating beyond the R ($d=0.0$) and M1 ($d=1.0$) phases at $V=0,1,2$\,eV in black, orange, and red, respectively. (a) Internal energy. (b) Band gap showing either a metallic, negative-gap, or direct-gap state. Vertical lines indicate the last structure with zero gap for each $V$ value. (c) Inter-atomic occupation difference between the short- and long-bonds ($\Delta n_\times^{a_{1g}}$).}
    \label{fig:distortion_energy_occ_gap}
\end{figure}

In \Cref{fig:distortion_energy_occ_gap}, we show the total energy, the band gap, and the difference of the $a_{1g}$ component of the inter-atomic occupation matrix between SB and LB, $\Delta n_\times^{a_{1g}}$, as a function of the distortion coordinate, $d$. We first note that for all studied $V$ values, we observe two distinct regimes, a metallic one (with zero band gap) for a small $d$, and an insulating one for larger distortion. We note that we also find a narrow intermediate regime, indicated by a negative band gap in \Cref{fig:distortion_energy_occ_gap}(b). Here, the bands are already separated everywhere in $k$-space, but the minimum of the higher-lying group of bands (which become completely empty in the insulating state) is still lower than the maximum of the lower-lying group of bands (which become completely filled in the insulating state).

For $V=0$ the system clearly exhibits two energy minima~[black curve in \Cref{fig:distortion_energy_occ_gap}(a)], one in the metallic regime for zero distortion and a second one, corresponding to a distorted insulating state around $d=1.25$. In agreement with previous literature~\cite{Eyert:2002, Lu_et_al:2020}, the minimum corresponding to the R phase is slightly lower in energy than that corresponding to the M1 phase, and this remains true also with our simplified structure. Additionally, consistent with the results of the structural relaxations above, the M1 energy minimum lies at a larger distortion than the experimentally reported structure, at about 0.20-0.25 above the $d=1$ distortion point.

At $V > 0$, the energy of the M1 phase minimum relative to the R phase minimum significantly lowers, becoming the global energy minimum~[orange and red curves in \Cref{fig:distortion_energy_occ_gap}(a)]. In fact, it becomes unclear whether R remains even locally stable. In contrast, introduction of a $V$ has negligible effects on the \emph{positions} of the energy minima, again in agreement with the relaxation results. At finite $V$ values, we see an opening of the band gap for a smaller distortion than at $V=0$~[orange and red curves compared to the black in \Cref{fig:distortion_energy_occ_gap}(b)], coinciding with the onset of the insulating M1 phase energy minimum.

Finally, we also note the behavior of $\Delta n_\times^{a_{1g}}$. We observe a linear increase in the metallic regime which becomes strongly nonlinear for $V > 0$ compared to the more gradual change at $V=0$. This strong increase in the inter-atomic occupation leads to an abrupt change in the curvature of the total energy as a function of $d$ around the point where the negative gap opens up (note the vertical lines in \Cref{fig:distortion_energy_occ_gap}). The inter-atomic occupation difference plotted here also resembles the evolution of the electronic order parameter as function of structural distortion in other systems exhibiting coupled structural and electronic transitions \cite{Peil_et_al:2019, Georgescu/Millis:2022, Carta/Ederer:2022}, suggesting that the $\Delta n_\times^{a_{1g}}$ plotted here is indeed a suitable order parameter for describing the electronic dimerization in VO$_2$.

These results corroborate our findings described in the previous section, that applying a $V$ in the way presented here improves the overall description of VO$_2$, in particular with respect to the relative energies of the two phases and the presence of a band gap in the M1 phase, consistent with the experimental observations. 


\section{Summary and Outlook}

In summary, we showed that the DFT$+V$ method can provide an improvement over plain DFT in the description of VO$_2$, at low computationally cost and without introducing a bias by pre-selecting specific vanadium-vanadium pairs as dimers.

Applying the $V$ correction enhances the $a_{1g}$ bonding-antibonding splitting resulting from the vanadium-vanadium dimerization in VO$_2$, which supports the formation of an insulating band gap and also energetically stabilizes the M1 relative to the R phase. In this context the $V$ should be viewed as an empirical self energy correction, stemming from the on-site electron-electron interaction, rather than as an inter-atomic Coulomb interaction. Within a simple tight-binding model, the effect of the inter-atomic $V$ can be understood as a renormalization of the hopping amplitude between vanadium-sites, which also depends on the degree of inter-atomic hybridization represented by the inter-atomic occupation $n_\times$. 

Applying a $V$ on all vanadium-vanadium pairs along the $c$ direction in both the R and M1 phases, we obtain both the correct insulating or metallic behavior and the correct energetic ordering between these two phases, thus removing two of the main deficiencies of plain DFT applied to VO$_2$. Our approach hence provides an unbiased way of treating VO$_2$ without requiring a pre-patterning of vanadium-vanadium pairs. We thus expect this DFT-based method to be an ideal low-cost choice for future computational VO$_2$ studies that require a very large number of calculations (e.g., scanning of different strain states) or calculations that necessitate larger system sizes, such as doped VO$_2$. 

From our calculations, a value of  $V=2$\,eV appears the most appropriate for M1 VO$_2$, yielding the experimental band gap of around 0.6\,eV \cite{Ladd/Paul:1969} at the global energy minimum corresponding to a distortion of $d=1.25$. Independent of the specific value of $V$ (and in particular also for $V=0$), the energy minimum corresponding to the M1 phase is obtained for larger distortion (by about 20-25\,\%), i.e., stronger structural dimerization, compared to the experimentally reported M1 structure of Ref.~\cite{Longo_et_al:1970}. This could point towards a general deficiency of DFT-based methods, towards a strong temperature dependence of the dimerization (since the experimental structure was measured around room temperature whereas the DFT-based results correspond to zero temperature), or towards other effects (not included in the calculations) affecting the measured distortion amplitude. Furthermore, we note that the DFT-relaxed structure  indeed has a very small, but nonzero band gap, even for $V=0$. Although this is in principle already known from previous works~\cite{Wentzcovitch/Schulz/Allen:1994, Mohebbi_et_al:2023}, it does not appear to be well-acknowledged in the literature.

Finally, we point out the similarity of our results to our previous work using DFT+DMFT with bond-centered orbitals to describe VO$_2$~\cite{Mlkvik_et_al:2024}, with the energy landscape of the distortion between the R and M1 phases being in particularly good agreement. This further highlights the importance of inter-atomic effects in the physics of VO$_2$.

\section*{Acknowledgments}
This work was supported by ETH Z\"{u}rich and through the Swiss National Science Foundation (Grant No.~209454). Calculations were performed on the ETH Z\"{u}rich Euler cluster.

\bibliography{dftv}

\end{document}